\documentclass[preprint,12pt]{elsarticle}

\usepackage{amsmath,amssymb,amsthm} 
\usepackage{comment}
\usepackage{algorithm}
\usepackage{algorithmic}
\usepackage{verbatim} 
\usepackage[normalem]{ulem}
\usepackage{graphicx, subcaption}
\usepackage{tabularx}
\usepackage{url}

\journal{}

\begin{document}

\begin{frontmatter}

\title{Try Before You Buy: A practical data purchasing algorithm for real-world data marketplaces}

\author[affiliation1]{Santiago Andrés Azcoitia}
\author[affiliation2]{Nikolaos Laoutaris}
\address[affiliation1]{IMDEA Networks Institute, Univ. Carlos III, Leganes (Madrid) Spain}
\address[affiliation2]{IMDEA Networks Institute, Leganes (Madrid) Spain}

\date{November 2020}


\begin{abstract}
Data trading is becoming increasingly popular, as evident by the appearance of scores of Data Marketplaces (DMs) in the last few years. Pricing digital assets is particularly complex since, unlike physical assets, digital ones can be replicated at zero cost, stored, and transmitted almost for free, etc. In most DMs, data sellers are invited to indicate a price, together with a description of their datasets. For data buyers, however, deciding  whether paying the requested price makes sense, can only be done after having used the data with their AI/ML algorithms. Theoretical works have analysed the problem of which datasets to buy, and at what price, in the context of full information models, in which the performance of algorithms over any of the $O(2^N)$ possible \emph{subsets} of $N$ datasets is known \textit{a priori}, together with the value functions of buyers. Such information is, however, difficult to compute, let alone be made public in the context of real-world DMs.

In this paper, we show that if a DM provides to potential buyers a measure of the performance of their AI/ML algorithm on \emph{individual} datasets, then they can select which datasets to buy with an efficacy that approximates that of a complete information model. We call the resulting algorithm \emph{Try Before You Buy} (TBYB) and demonstrate over synthetic and real-world datasets how TBYB can lead to near optimal buying performance with only $O(N)$ instead of $O(2^N)$ information released by a marketplace.  
\end{abstract}


\end{frontmatter}

\section{Introduction}
\label{sect:Introduction}

Data is considered a key production factor, comparable in importance to labour, capital, and infrastructure. Companies are often in need of data they do not possess, or cannot collect directly. Therefore, general purpose\footnote{See for example DAWEX, Azure Data Catalog, or AWS Data Exchange} and domain specific\footnote{See for example, Openprise, Lotame PDX (marketing), Qlik (business intelligence), or Battlefin (investment information)} data marketplaces (DMs) have appeared with the purpose of building a business of mediation between data selling and data buying \emph{companies}. Leading data management platforms and innovative startups\footnote{See for example Snowflake, Cognite, Carto, and Openprise} are also introducing marketplace functionalities into their products. Finally, personal information management systems (PIMS) have answered the call of recent legislative developments in personal data protection by offering data control, portability, and monetization services for \emph{individuals}.

Designing and building a successful DM calls for solving a plethora of technology, business, and economics challenges in the context of a complex two-sided market (see \cite{Armstrong06, Rochet06, Rysman09} for an exposition). According to our survey of more than 75 real-world data marketplaces, the most common bootstrapping strategy is for a DM to spend effort and money to attract a sufficient set of data sellers, and then try to convince as many buyers as possible to start purchasing these datasets. Therein lie two fundamental problems: (1) the \emph{dataset pricing problem} for data sellers, and (2) the \emph{dataset purchasing problem} for data buyers. 

Recent theoretical work on the intersection between computer science and economics has looked into those problems, and has proposed solution concepts and algorithms for them \cite{Agarwal19, Chen19, Chawla19, Koutris15}. For data sellers, selecting efficient prices requires knowing the level of competition with other data sellers, the willingness to pay of buyers, potential customer lock-ins and other information that affects prices in digital and non-digital markets. For buyers, the problem of selecting which datasets to buy (problem (2)), given the prices set by sellers (problem (1)), can be further broken down into 2 interrelated subproblems: (2.a) compute how useful these datasets will be to their AI/ML algorithms, something that can be captured by various accuracy metrics, and (2.b) compute how such accuracy can be converted into monetary gains (via e.g., improved sales, acquisition of new customers, retention of existing ones, etc.). Sub-problem (2.b) is probably the easiest of the two challenges faced by buyers, since most companies are able to gather historical data about the impact of things like recommendation quality on actual sales \cite{Brovman16}. Subproblem (2.a), on the other hand, is inherently more challenging, since buyers need to have access to the data before they can compute their value for their AI/ML task, but such access is only granted \emph{after} a dataset purchase has taken place -- a chicken and egg problem essentially. (2.a) is further exacerbated when the buyer can/has to buy more than one dataset in order to improve the accuracy of its AI/ML algorithm. With $N$ available datasets, a buyer has $O(2^N)$ data purchase options each one with a cost equal to the sum of individual dataset prices and a value defined by the maximum accuracy of AI/ML algorithm operating over the aggregate data.   

In theoretical works, the value of any subset of datasets for a data buyer is considered as known \textit{a priori} \cite{Shen16}. In reality, however, things are completely different \cite{Hubert18}. In almost all the 75 DMs that we have surveyed, data sellers provide only a description of their datasets, a price, sometimes an outdated sample, and buyers have to make purchase decision with that information alone. Few of these DM (e.g., Dawex, Airbloc, Wibson or Databroker) also allow buyers to make offers (bids) for data when sellers do not indicate a fixed price, or if they are willing to pay something below the asking price. This case suffers as well from the fundamental problem (2.a) of not knowing the value of a dataset before purchasing it. 
\vspace{2pt}

\noindent \textbf{Our contribution:} In this paper we show how to solve the \emph{dataset purchasing problem} for data buyers in a way that approximates the efficiency of an optimal full information solution, but in a way that is implementable in practice with real-world DMs. Our main contribution is a family of dataset purchase algorithms that we call "Try Before You Buy" (or TBYB) that allow data buyers to identify the best datasets to buy with only $O(N)$ information about the accuracy of AI/ML algorithms on individual datasets, instead of $O(2^N)$ information used by an optimal strategy using full information. Effectively, TBYB needs to know only the accuracy of an AI/ML on \emph{individual} datasets, and with this information it can approximate the optimal \emph{combination} of datasets that maximizes the profit of the buyer, i.e., the difference between the value extracted from the datasets minus the cost of purchasing them. The accuracy of individual datasets can either be precomputed by the DM or the data sellers, and be made available as part of the dataset description (e.g., for some common AI/ML algorithms). Another alternative is for the DM to use recently developed ``sandboxed'' environments that allow data buyers to experiment versions of the data without being able to copy or extract them (hence the ``Try'' part on the algorithm's name; Otonomo, Advaneo, Caruso or Battlefin are examples of marketplaces that implement such functionality). Overall, with TBYB our objective is to increase the efficiency of buying datasets online from DMs. We believe that this is key for allowing both DMs and the data offer side to grow, as well.
\vspace{2pt}

\noindent \textbf{Our findings:} We compare the performance of TBYB against several heuristics that do not use information about the value of a datasets for the particular AI/ML task at hand, as well as against an optimal solution that uses full information. We start with a synthetic evaluation and then validate our conclusions using real-world spatio-temporal data and a use case in predicting demand for taxi rides in metropolitan areas \cite{Andres20}. Our findings are as follows:
\begin{itemize}
    \item TBYB remains close to the optimal for a wide range of parameters, whereas its performance gap against the heuristics increases with the catalog size.
    \item TBYB is almost optimal when buying more data is yielding a progressively diminishing return in value for the buyer (i.e., when the value function of the buyer is concave). With convex function, it becomes increasingly difficult for TBYB to match the optimal performance. It's performance gap with the heuristics, however, is maintained. 
    \item When the asking price of datasets does not correlate with their actual value for the buyer, the performance advantage of TBYB over the heuristic becomes maximal. When the pricing of data follows their value for buyers, the performance of TBYB is still superior but the gap with the heuristics becomes smaller. 
\end{itemize}

Overall, our work demonstrates that near optimal dataset purchasing is realistic in practice and that it could be implemented relatively easy by real-world data marketplaces. 

\section{Marketplace Model \& Definitions}
\label{sect:Definitions}

Existing DMs typically list the datasets that they make available and provide for each one a description and a price. In our case we will assume that the DM also provides for each dataset its \emph{accuracy} over a range of common AI/ML tasks. This list cannot/need not be exhaustive, nor it needs to capture all the specificities of the particular AI/ML algorithm that the buyer intends to use. The intention is merely to provide the buyer with a hint, even an \emph{approximate} one, regarding the accuracy that he should expect if he buys a particular dataset.

If, on the other hand, a buyer would like to know before buying a dataset the \emph{exact} performance of his algorithm on the data, then the following two options exist. The buyer could submit a description of the task for which he needs data, so that the DM returns a list of candidate sellers, or, alternatively, she could just go over the data catalog and select the best candidates manually. In both cases, the DM can provide a sandboxed environment in which the buyer can submit his \emph{exact} algorithm and get an \emph{exact} answer in terms of the achieved accuracy over each candidate dataset, without being able to see/copy the raw data.

\begin{figure}
\centering
\includegraphics[width=0.8\textwidth]{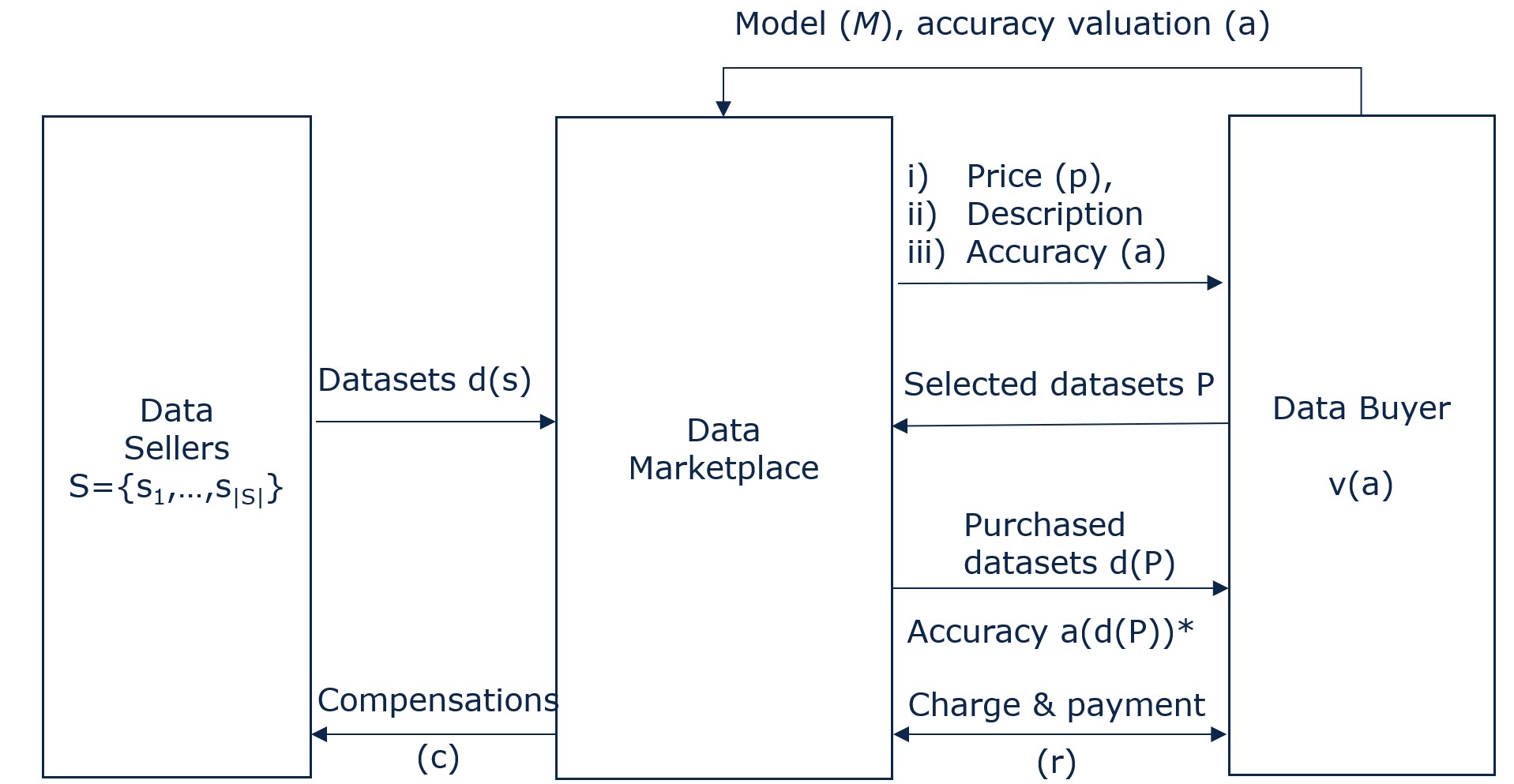}
\caption{Reference DM model}
\label{Fig:model}
\end{figure}

To model any of the above cases (see figure \ref{Fig:model}), we will denote by $\mathcal{S}$ the set of suitable sellers for the AI/ML task of a particular buyer. We will denote by $d(s)$ the dataset offered by seller $s \in \mathcal{S}$, by $p(s)$ its \emph{price}, and by $a(d(s))$ the \emph{accuracy} that the buyer's AI/ML task can achieve if trained by $d(s)$. Similarly, for a subset of the sellers $S\in \mathcal{S}$, we will denote by $d(S)$ their aggregated dataset, and by $a(d(S))$ the maximum accuracy that can be achieved using all or a subset of the data in $d(S)$. We will also introduce the \emph{value function} $v(a)$ of the buyer, that indicates the (monetary) value that the buyer can achieve when his AI/ML algorithm gets to an accuracy of $a$. In Sect.~\ref{subsec:TheorSensitivityMUP} we will look at both concave and convex $v(\cdot)$ functions.

When datasets are sequentially bought, we will append a subscript to identify the round to the notation we already defined. Hence, $S_n \subset \mathcal{S}$ will refer to the set of eligible datasets in round $n$. We will denote by $P_n$ the set of data already under the buyer's control in round $n$. As a result, she is able to achieve an accuracy equal to $a_n=max_{S'\in 2^{P_n}} a(S')$ and a value $v_n=v(a_n)$.

Buyers will always look forward to optimizing their profit, hence a greedy buyer will decide to purchase a dataset $d(s)$, if its marginal value exceeds its cost, i.e. $v(a(d(s|P))) \geq {p(s)}$. In the event that such a value is unknown, the buyer must estimate it, and assume the risk that the purchase may not fulfil expectations.

\section{Data Purchase Strategies}
\label{sect:Purchase algorithms}

In this section we will present a series of data purchase strategies that cover the spectrum from full information, i.e., knowing the accuracy over any subset of the available data, to having no information about accuracy, as is currently the case in most DMs. In between the two extremes, lies our proposed algorithm called TBYB, that runs only on accuracy information of individual datasets.

\subsection{Optimal purchase under full information}
\label{subsec:fullinformation}
In this case, the buyer knows $a(d(S))$ for any subset $S\subseteq 2^\mathcal{S}$. This allows for an optimal purchase $\mathcal{S}^\star$ that maximizes the profit, i.e., the difference between the value that the buyer extracts from the data, and the cost paid to purchase them: 

\begin{equation}
\label{eq:optimalfull}
    \mathcal{S}^\star =  arg\,max_{S\in 2^\mathcal{S}}  \left(v(a(d(S)))-\sum_{s \in S} p(s) \right),
\end{equation}

subject to $v(a(d(\mathcal{S^\star}))) \geq \sum_{s\in \mathcal{S}^\star} p(s)$.

Such a full information scenario is optimal from a buyer's perspective, but not scalable nor practical: a DM would need to compute the accuracy of each AI/ML algorithm over $2^{|\mathcal{S}|}$ combinations of eligible datasets.

\subsection{Try Before You Buy (TBYB)}
\label{subsec:TBYB}
For our proposal, we assume that the DM provides the buyer with the accuracy of her algorithm on individual datasets, but not on combinations of them. The algorithm is sequential and greedy in nature, and can run for up to $|\mathcal{S}|$ iterations. We will consider two versions.

\subsubsection{Stand-alone version - S-TBYB}
\label{subsec:S-TBYB}
The marketplace provides $a(d(s))$ on all $s\in\mathcal{S}$. Then the algorithm starts buying datasets in descending order of \emph{expected profit} until a stopping condition is reached. For the first dataset, the profit is not expected but exact, so the best dataset is bought provided $v(a(d(s)))- p(s) \geq -\lambda \cdot v(a^\star)$, where:
\begin{enumerate}
    \item the best accuracy $a^\star \leq 1$ that can be delivered by the marketplace. Either the data marketplace provides data buyers with this information, or the buyer makes her best guess, and
    \item the risk parameter $\lambda$ models the maximum relative admissible loss the buyer is willing to assume in each operation. The risk assumed every round will be bounded by $\lambda$ times the potential value of the sourcing operation which, in round n, is equal to $v(a^\star) - v_n$. For example, $\lambda = 0.1$ means that the buyer will buy a new dataset $s$ if its price is lower than the marginal value she expects to get plus $10\%$ of the maximum value that she could add by buying new data.
\end{enumerate}

In some sourcing problems, the marginal value of new data increases as more information is bought. In such a setting, buyers may be required to assume some temporary losses when acquiring the first datasets, in the hope that they provide additional accuracy, and become profitable when fused together with other data.

\paragraph{n-th iteration}The buyer will proceed as follows:
\begin{itemize}
        \item Identify the best possible dataset $s^\star \in S_n$ such that: 
        \begin{equation}
            s^\star =  arg\,max_{s\in S_n}  \left( v(a(d(s)))- p(s) \right)
        \end{equation}
        \item Purchase $s^\star$ if its estimated marginal value exceeds its price, and a risk threshold that depends on the remaining value she expects to get out of the operation, i.e., if $v(E\{a(s^\star \cup S_n)\}) - v_n - p(s^\star) \geq -\lambda \cdot (v(a^\star) - v_n)$
        \item If the buy condition is met then, $s^\star$ is added to the set of controlled datasets: $P_{n+1} = P_n \cup d(s^\star)$ and the next round starts
        \item else if no dataset in $S$ meets this requirement, then the process stops
\end{itemize}

To estimate $E\{a(s^\star \cup S_n)\}$ the buyer could use the following information:

\begin{enumerate}
    \item The price and accuracy pairs $<p(s),v(s)>$ for all individual datasets $s\in S$
    \item The accuracy of every possible combination of already purchased datasets, i.e., the $v(S')$ for all $S' \subseteq 2^{P_n}$
\end{enumerate}

This estimation must be tailored to each specific problem, and it turns out to be non-trivial. We estimate the relative added accuracy of $s^\star$ by multiplying its individual accuracy $a(s^\star)$, and the ratio of the marginal contribution and the individual accuracy of the last purchased dataset: 
\begin{equation}
\label{eq:ExpectedAccuracyEstimation}
    E\{a(s^\star \cup S_n)\} = \frac{a_n - a_{n-1}}{a(P_n - P_{n-1})} \cdot a(s^\star) \cdot (a^\star - a_n).
\end{equation}

\subsubsection{Assisted version - A-TBYB}
\label{subsec:A-TBYB}
In this case, we will assume the buyer is allowed to ask the marketplace \textit{every round} for the marginal accuracy of any eligible datasets given the data she already owns.

\paragraph{n-th iteration}The purchase process will be the following:
\begin{enumerate}
    \item Ask the marketplace for complementary datasets $S_n \subseteq \mathcal{S}$, and $a(d(s)|P_n),$ $\forall s \in S_n$ given the task $(\mathcal{M}, a)$ and $P_n$
    \item If $S_n \neq \emptyset$: \begin{itemize}
        \item Identify the best possible dataset $s^\star \in S_n$ such that:
        \begin{equation}
            s^\star =  arg\,max_{s\in S_n}  \left( v(a(d(s \cup P_n)))- p(s) \right)
        \end{equation}
        \item Buy provided $v(a(d(P_n \cup d(s^\star)))) - v_n - p(s^\star) \geq -\lambda \cdot (v(a^\star) - v(a(P_n)))$
        \item If the buy condition is met then, $s^\star$ is added to the set of controlled datasets: $P_{n+1} = P_n \cup d(s^\star)$ and the next round starts
        \item Else if the buy condition is not met, then the process stops
    \end{itemize}
\end{enumerate}

As a result, if the marketplace is asked to compute the marginal accuracy for every remaining dataset every round, the model will be processed a maximum if $\sum^{r-1}_{i=0}{|S|-i}$ times for $r$ rounds. To prevent abuses from buyers, a marketplace implementing this solution could set up a maximum limit of trials for a certain task. Such a limit may be updated as the buyer purchases data.

\subsection{Buying without trying}
\label{subsec:BuyingWithoutTrying}
\subsubsection{Volume-based purchasing}
\label{subsec:volumePurchasing}
Most commercial marketplaces provide buyers with a description of datasets, their metadata, source, procedure used to collect them, etc. Oftentimes, the volume of data in a particular dataset (e.g. nº observations or samples) is used as the deciding figure of merit for choosing among different offers. Let $vol(s)$ denote the volume of dataset $s$, used as merit figure by the following volume-based purchasing heuristic:

\paragraph{n-th iteration}We will assume that a greedy buyer would select the dataset $s^\star \in S_n$ with the highest $vol(s) / p(s)$ ratio every round. However, it is not possible to know the accuracy it will yield to the specific problem. We assume a conservative condition for the algorithm to decide to purchase $s^\star$, specifically:

\begin{equation}
    p(s^*) \leq -\lambda \cdot (v(a^\star) - v_n),
\end{equation} 

which assumes that even in the worst case, where the purchase does not improve $(\mathcal{M},a)$ accuracy at all, the maximum relative admissible loss is not exceeded in the operation.

\subsubsection{Price-based purchasing}
\label{subsec:PriceBasedPurchasing}
It may happen that the marketplace just publishes the list of suitable datasets $\mathcal{S}$, and their prices. This setting resembles real situations where information about data offer is insufficient or misleading to the buyer's purposes. We assume such a buyer would randomly select among datasets whose price is lower than their maximum relative admissible loss.

\paragraph{n-th iteration}The buyer will randomly select one of the datasets $S_n \subseteq \mathcal{S}$ such that, $\forall s \in S_n, p(s) \leq - \lambda \cdot (v(a^\star) - v_n)$. If $S_n = \emptyset$ then the process stops.

\section{Performance evaluation with synthetic data}
\label{sect:TheoreticalEvaluation}
We will use synthetic data to evaluate the performance of the different purchase strategies of Section \ref{sect:Purchase algorithms} across a wide range of parameters. Our synthetic model is easy to reproduce, captures a wide range of parameters, and allows us to extract useful insights about the relative performance of different data purchase strategies. As we will show later in Sect.~\ref{sect:ValidationWithData}, our conclusions from this section are also validated by results with real data.

\subsection{Synthetic model description}
\label{subsec:TheoreticalModel}

To simplify the evaluation, we will assume that the value for the data buyer will be equal to the accuracy $a$. Hence, the maximum value that a buyer can extract from data is equal to 1, which occurs when the accuracy of its AI/ML algorithm trained on the purchased data, becomes 1 (i.e., 100\%). We will denote as Total Cost of Data (TCOD), the cost of buying all the available datasets in $\mathcal{S}$, i.e., $TCOD=\sum_{s\in \mathcal{S}} p(s)$. Therefore, when $TCOD<1$, then the buyer is guaranteed to make a profit, independently of the data purchase strategy used. In the more interesting case of $TCOD\geq 1$, the buyer needs to select which datasets to buy carefully, to avoid ending up with a loss.      

Having equated value with accuracy, we will also need to connect the datasets bought, with the achieved accuracy. For a data buyer that buys datasets $S\subseteq \mathcal{S}$, the value will be given by the following expression:

\begin{equation}
    v(a(d(S))) = a(d(S)) = \left( \frac{\sum_{s_i \in S}{DI^{i}}} {\sum_{s_i \in \mathcal{S}}{DI^{i}}} \right) ^{MUP},
\end{equation}

where:
\begin{itemize}
    \item \textbf{MUP is the Marginal Utility Profile parameter} that controls the concaveness/convexity of $v$ from 0 to 1 as more data is bought. When $MUP<1$, then buying additional datasets will have a decreasing marginal utility in terms of accuracy, and hence value for the buyer, both of which will be concave with respect to the amount of data bought. On the other hand, with $MUP>1$, the marginal contribution of new data sources will be increasing as more datasets are bought, making $v(\cdot)$ and $a(\cdot)$ convex. Finally, $MUP=1$ means that all datasets yield the same accuracy if they are bought first, and the same incremental change if they are bought second, and so forth.
    \item \textbf{DI is the Data Interchangeability parameter} that controls the relative importance of different datasets in $\mathcal{S}$. Setting DI equal to 1, amounts to making all datasets fully interchangeable. Therefore, in this case, it only matters how many datasets are bought, but not which ones. For $DI>1$ and $MUP=1$, dataset $s_i$ becomes DI times more important than dataset $s_{i-1}$, $1\leq i\leq |\mathcal{S}|$. Effectively, for $DI\neq 1$ what matters is not only how many datasets are bought, but also which ones.   
\end{itemize}

The last element of our synthetic model has to do with how we set the prices of individual datasets. We will consider the following pricing schemes:

\begin{itemize}
    \item \textbf{All datasets having the same price}, i.e., $p(s)=$TCOD$/|\mathcal{S}|$, $\forall s\in \mathcal{S}$.
    \item \textbf{Datasets having random prices} drawn from a uniform distribution in $[0,1]$ and scaled to add up to TCOD. 
    \item \textbf{Datasets having a price that reflects their importance} captured by their Shapley value within a coalition of $\mathcal{S}$ datasets that achieve a total value equal to $a(d(\mathcal{S}))$ (see works such as \cite{Ghorbani19, Paraschiv19} for a justification and explanation about hot to use the Shapley value with aggregate datasets).
\end{itemize}

\subsection{Results for different marginal utility profiles}
\label{subsec:TheorSensitivityMUP}

\begin{figure}
\centering
\includegraphics[width=\textwidth]{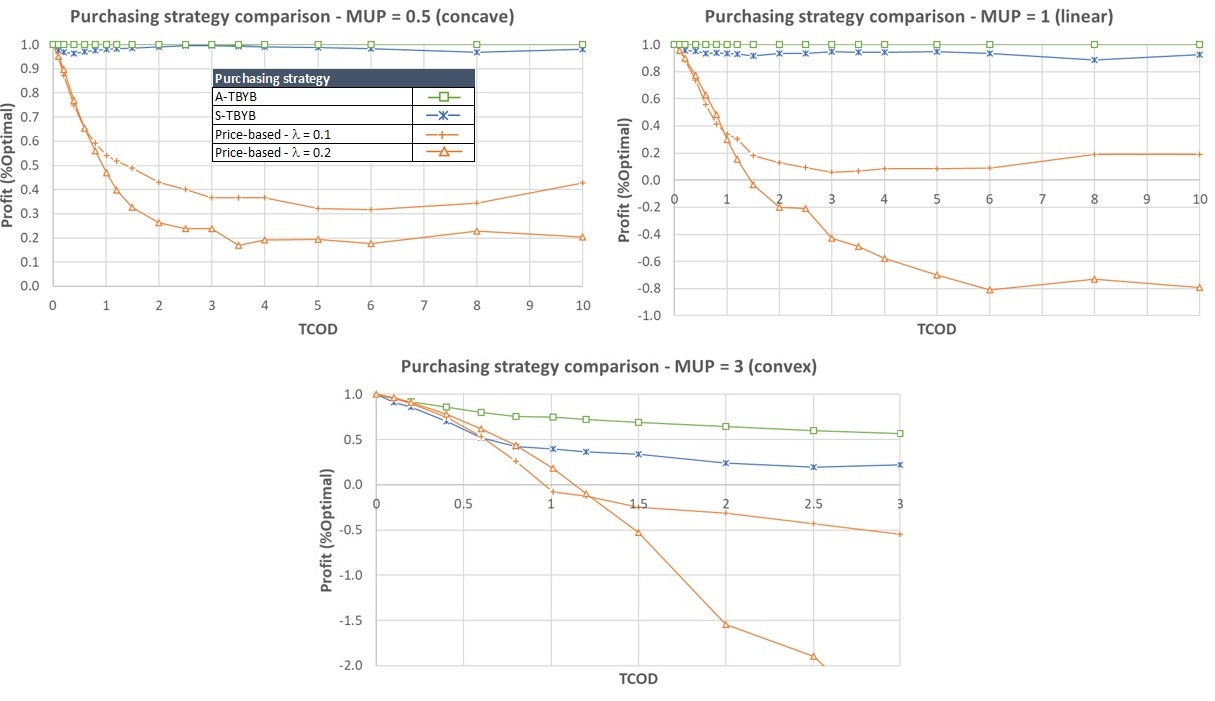}
\caption{Profit (\% of the optimal) vs. TCOD for different MUP and value-unrelated prices}
\label{Fig:MUPSensitivity Random Price}
\end{figure}

\begin{figure}
\centering
\includegraphics[width=\textwidth]{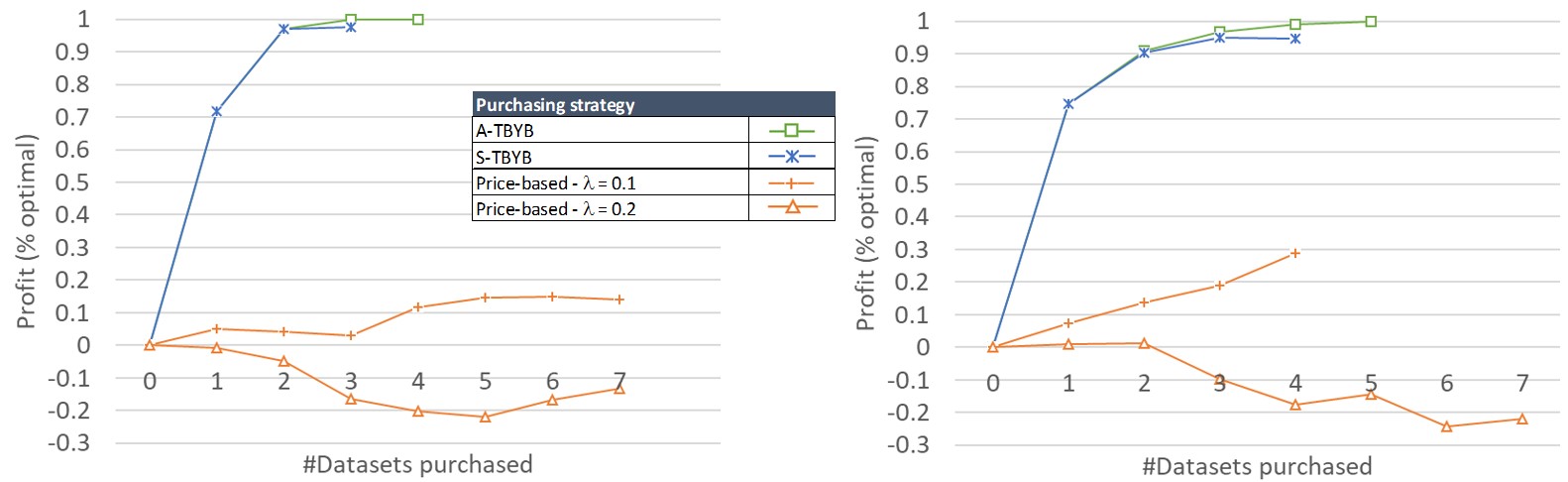}
\caption{Purchase sequences for MUP = 1 and TCOD = 1.5 (left), and 3 (right)}
\label{Fig:PurchasingSequencesTheoretical}
\end{figure}

Next we will compare optimal purchasing under complete information, with TBYB, and the value-agnostic heuristics across a range of parameter values for capturing things like the total cost of datasets (TCOD), the marginal utility of buying more datasets (MUP), their relative value (DI), and their relative price. Our main evaluation metric will be the profit for a data buyer, i.e., the value extracted from the data minus the cost paid to obtain them. As stated in the introduction, guaranteeing that buyers obtain a healthy profit from data is vital for bootstrapping the nascent data marketplace sector. In the future we will also examine seller side profits and social welfare. Of course doing the latter makes sense for already bootstrapped markets. It also requires modeling complex market dynamics such as competition, dynamic pricing, etc. that go beyond the scope of the current work. Whenever randomization is used, e.g., in pricing, or in some of the value-agnostic heuristics, we report average values from 50 executions.   

The first parameters that we examine are TCOD and MUP, assuming some datasets are more important than others ($DI = 2$). Obviously, as data become more expensive (higher TCOD), all the strategies, including the optimal one, yield a smaller profit for buyers. What we are interested to study, therefore, is the relative performance of different strategies under different TCODs and MUPs. Figure \ref{Fig:MUPSensitivity Random Price} shows that A-TBYB matches the optimal purchasing for both concave (MUP=0.5, left subplot) and linear (MUP=1, middle subplot) value profiles, across the entire range of TCOD values. For convex value profiles (MUP=3, right subplot), A-TBYB ceases to be optimal, but remains the best performing strategy. 

What is even more interesting, is the performance of the much simpler to implement S-TBYB, which stays above 90\% of the optimal for concave and linear MUPs, when the value agnostic heuristics drops below 50\% with TCOD>1 and even leads to losses (see MUP=1 results). Under convex value profiles (MUP=3), all strategies yield a lower performance, since reaching a higher accuracy (and therefore buyer value) requires buying more datasets, which, in turn, eats away the profit margins for buyers. Even in these cases, S-TBYB yields a profit and avoids loss. 

To explain why TBYB outperforms the value unaware heuristics, we plot in Fig.~\ref{Fig:PurchasingSequencesTheoretical} a series of ``\textit{purchase sequences}'', demonstrating the evolution of profit with the number of datasets purchased (by different algorithms). As shown in the plot, TBYB algorithms buy both the most valuable datasets (they achieve higher profits from the first round), and the right number of them (they stop buying before profits decrease). On the other hand, value unaware heuristics overbuy, and randomly select datasets they can afford according to their risk appetite, which generally lead to lower profits, or even loses especially for risk-prone buyers.

\subsection{The effect of data interchangeability}
\label{subsec:TheorSensitivityDI}
\begin{figure}
\centering
\includegraphics[width=\textwidth]{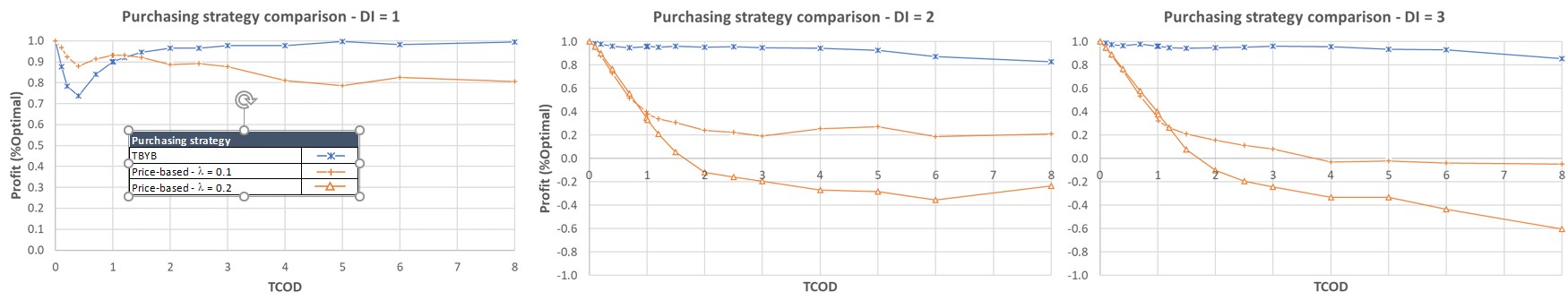}
\caption{Profit for purchase algorithms (MUP = 1, linear) with different DI values}
\label{Fig:DISensitivity}
\end{figure}

To find out how TBYB is affected by the interchangeability of datasets, we have run a set of simulations for different values of the parameter DI. Figure~\ref{Fig:DISensitivity} shows three different plots of the relative profit of different purchase algorithms for different DI values under MUP=1. The subplot on the left depicts results for perfectly interchangeable datasets (DI = 1), whereas the next two show cases of datasets that are increasingly less interchangeable (DI=2 and DI=3). 

These plots show that the performance benefits of TBYB over the heuristics increase when different datasets have different value in terms of the accuracy they can achieve, both alone, as well as combined with other datasets. This happens, of course, because the advantage of knowing the value of data before buying them, gets diminished when datasets are almost interchangeable. In reality, as we will show in the next section, real world datasets are not interchangeable, which means that value unaware heuristics will not be able to match the performance of TBYB.


\subsection{The effect of data pricing}
\label{subsec:TheorSensitivityPricing}
In this section, we look at the role of dataset pricing on the performance of TBYB. Our main interest is to see what happens when the price of a dataset is proportional to its value for an AI/ML algorithm, and when it is not. The former we create via the Shapley value method discussed in Sect.~\ref{subsec:TheoreticalModel}. The latter we model it in two ways: with datasets that have the same price but yield different accuracy, and with datasets that have randomly distributed prices and different accuracy. 

\begin{figure}
\centering
\includegraphics[width=\textwidth]{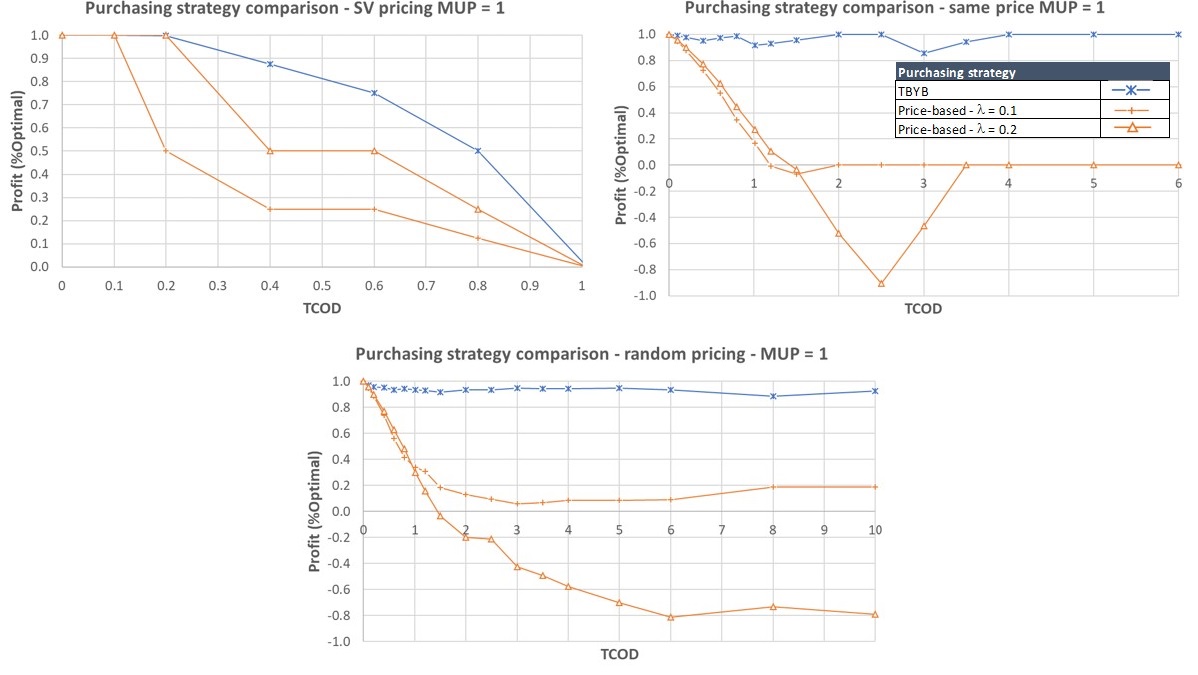}
\caption{Profit for different pricing methodologies (MUP = 1, DI = 2)}
\label{Fig:PricingSensitivity}
\end{figure}

Figure \ref{Fig:PricingSensitivity} shows the results of our purchase algorithms for different pricing models. In every case, A-TBYB matches the optimal. Pricing data based on their real value for AI/ML algorithms reduces the gap of S-TBYB vs. price-based purchasing, although S-TBYB still outperforms price-based purchasing for the same level of risk. Notice, however, that pricing data in accordance to their actual value for buyers, requires knowing the value function of each buyer, something that buyers, of course, have no incentive to disclose to sellers. Even if they did, different buyers may have different value functions, so, in general, it cannot be expected that the price of a dataset will follow its value for different buyers that may be using it with different AI/ML algorithms, and having different value functions.\footnote{Notice that to simplify our synthetic evaluation we have assumed that buyer value follows the accuracy achieved by each dataset. This, of course, need not apply in the real world, since different buyers may have radically different value functions that translate accuracy into monetary worth.}

\subsection{Summary}
\label{subsec:TheorConclusions}
Table \ref{tab:ParametersImpact} summarizes the impact of parameters in the performance gap between TBYB and price-based purchasing. In summary, TBYB, even in its simplest, stand-alone version, always outperforms the value unaware heuristics, especially in the most realistic scenarios involving high TCOD, concave value functions, non-interchangeable datasets, and pricing that does not follow value. In a good part of the parameter space TBYB is very close to the performance of optimal purchasing that uses full information. 

\begin{table}
  \caption{Impact of parameters on the gap between TBYB and price-based purchasing}
  \label{tab:ParametersImpact}
  \scriptsize
  \begin{tabular}{p{1.2cm}p{5cm}p{6cm}}
    \hline
    Parameter&Impact&Explanation\\
    \hline
    \texttt{TCOD}&The higher TCOD, the more valuable TBYB&More difficult for other strategies to find the right datasets in terms of price - value to buy \\
    \texttt{MUP}&The higher MUP, the more difficult to find the optimal. TBYB loses effectiveness but still outperforms other algorithms&TBYB buys more valuable datasets, minimizes temporary losses and limits risk for buyers, since it allows for a better estimation of expected marginal value of datasets\\
    \texttt{DI}&The less interchangeable datasets are, the more advantage of using TBYB&With perfectly interchangeable datasets, TBYB only improves the estimation of marginal utility as information increases \\
    \texttt{Pricing}&TBYB gap with price-based purchasing narrows when prices are not related to value&Price-based purchasing works better if value is embedded in price\\
    \hline
  \end{tabular}
\end{table}

\section{Validation with real data}
\label{sect:ValidationWithData}

The synthetic model of the previous section limits the ways in which two or more datasets may mix, and impact the accuracy of an AI/ML algorithm. It allows only for concave/convex mixing with equal (interchangeable, DI=1) or unequal (non-interchangeable, DI>1) contributions to accuracy from the different datasets. In reality, however, different datasets may mix in much more complex ways, that cannot be represented by any parameter setting of the above model. For example, a certain dataset $d_i$ can be very useful if combined with another dataset $d_j$, but not so useful if combined with others that individually yield the same accuracy as $d_j$ does. To verify our conclusions from the previous section, we tested the performance of different data purchase strategies using real spatio-temporal data, in a use case that involves forecasting demand for taxi rides in a city.

Furthermore, in this section we expand our performance evaluation by introducing a new data pricing scheme, and a new data purchase strategy:
\begin{enumerate}
    \item \textbf{Volume-based pricing}. In this case the price of a dataset becomes proportional to its volume. In our use case, volume will correspond to the number of  drivers in the company.
    \item \textbf{Volume-based purchasing}. We will test the performance of a new heuristic that seeks to purchase that largest possible dataset in terms of volume for a given price.
\end{enumerate}

According to an internal survey covering more than 75 companies in the data economy, pricing and purchasing data by volume is a commonly extended practice in data trading. Therefore we compare TBYB against those practices as well. 

\subsection{Use case description}
\label{subsec:ValidationUseCaseDescription}
We will assume that a data buyer is looking to purchase datasets for training a multiseasonal SARIMA forecasting model with the purpose of forecasting, at an  hourly timescale, the demand for taxi rides in different districts of Chicago City, for the weeks to come. To achieve this objective, a number of taxi companies will be assumed to be the data sellers, that release historical data on past trips that they have provided in the \textit{observation period} or $T_o$. Such datasets are publicly available, thanks to reporting obligations that such companies have to fulfil towards the local authorities \cite{TaxiTrips19}. The accuracy of the forecasting algorithm is quantified by how accurate the algorithm can predict real demand observed in a \textit{control period} or $T_c$. Our model is able to accommodate any sequence similarity metric in order to compare predicted vs. real demand in $T_c$. 

\subsection{Dataset description}
\label{subsec:ValidationDatasetDescription}
From the above-mentioned repository \cite{TaxiTrips19}, we have obtained 11.1 MM rides corresponding to the first 8 months of 2019. These rides are included in 15 datasets that correspond to the 15 largest taxi companies in the city (servicing 94\% of the total demand), plus a hypothetical 16th company that aggregates all the rides reported by the remaining smaller companies. These will be our 16 data sellers according to our problem formulation.

We computed the exact Shapley value of data from each company to the forecasting accuracy achieved by the multiseasonal SARIMA model in predicting the demand in the second half of April using taxi rides from the previous six weeks for training ($T_o = $ Mar. 4th - Apr. 14th and $T_c = $ Apr. 15th - 28th). We deliberately chose to predict the taxi demand of a medium size district of the city (community area 11, Jefferson Park), where data from several companies is needed in order to achieve a good prediction accuracy. As a result, the Shapley values are very different for each source (standard deviation = 76\% of the average). Moreover, these were found to be weakly correlated with the number of licenses of each company ($R^2 = 0.54397$), because big companies usually concentrate in other areas of the city.

Unlike what happened in the theoretical use case, the maximum accuracy the marketplace can give using all the information is $a^\star = 0.896294$ in this case. As in the synthetic case, we will assume that the economic value of a prediction is equal to its accuracy.

\subsection{Empirical results}
\label{subsec:ValidationResults}
\begin{figure}
\centering
\includegraphics[width=\textwidth]{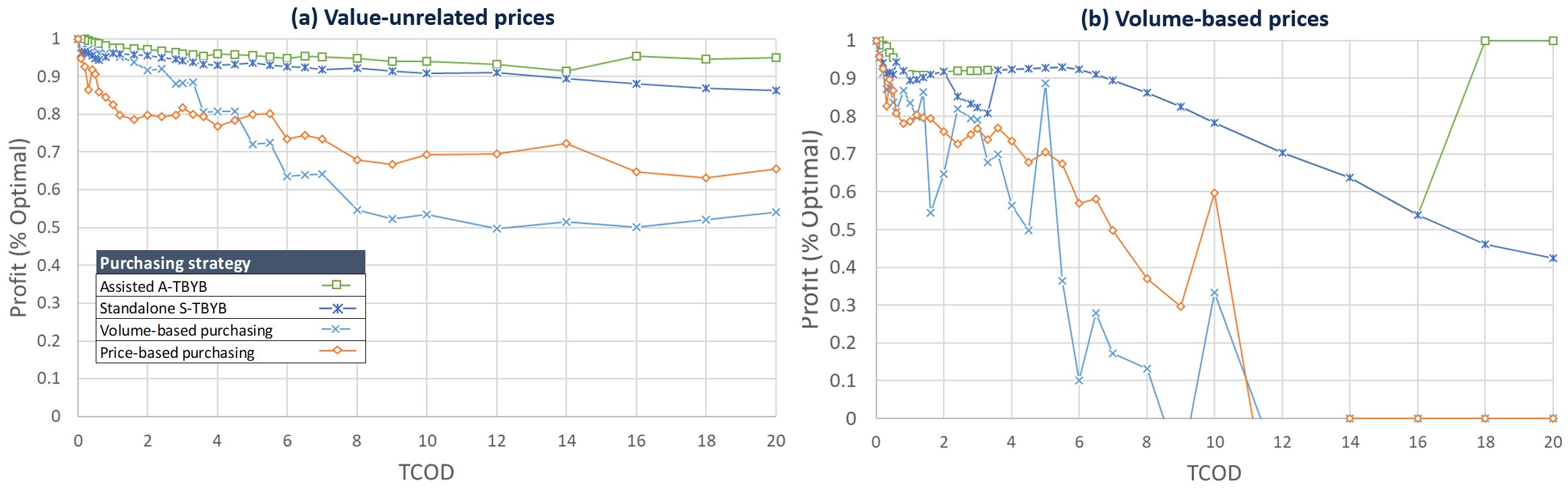}
\caption{Profit for different purchase strategies when prices are (a) unrelated to value, and (b) related to volume}
\label{Fig:ResultsChicagoD11}
\end{figure}

We have simulated all purchase algorithms for different TCOD, pricing models, and $\lambda$ parameters. Figure~\ref{Fig:ResultsChicagoD11} (a) shows that both A-TBYB and S-TBYB achieve above 90\% of the optimal buyer's profit under value-unrelated dataset pricing. The results are in line with the ones we obtained using synthetic data.

Regarding volume-based purchasing, it proved to outperform price-based purchasing, but only when TCOD is low (< 5). This is because value and volume are not tightly correlated in this case, hence buying by volume does not lead necessarily to higher accuracy.

Looking at Fig.~\ref{Fig:ResultsChicagoD11} (b) we see the corresponding results under volume-based prices. In this case, profit reduces faster than in the case of value-unrelated prices as TCOD grows, since valuable datasets are set higher prices. Still TBYB outperforms buying without trying algorithms, since it selects cheaper and more valuable datasets.

\begin{figure}
\centering
\includegraphics[width=0.5\textwidth]{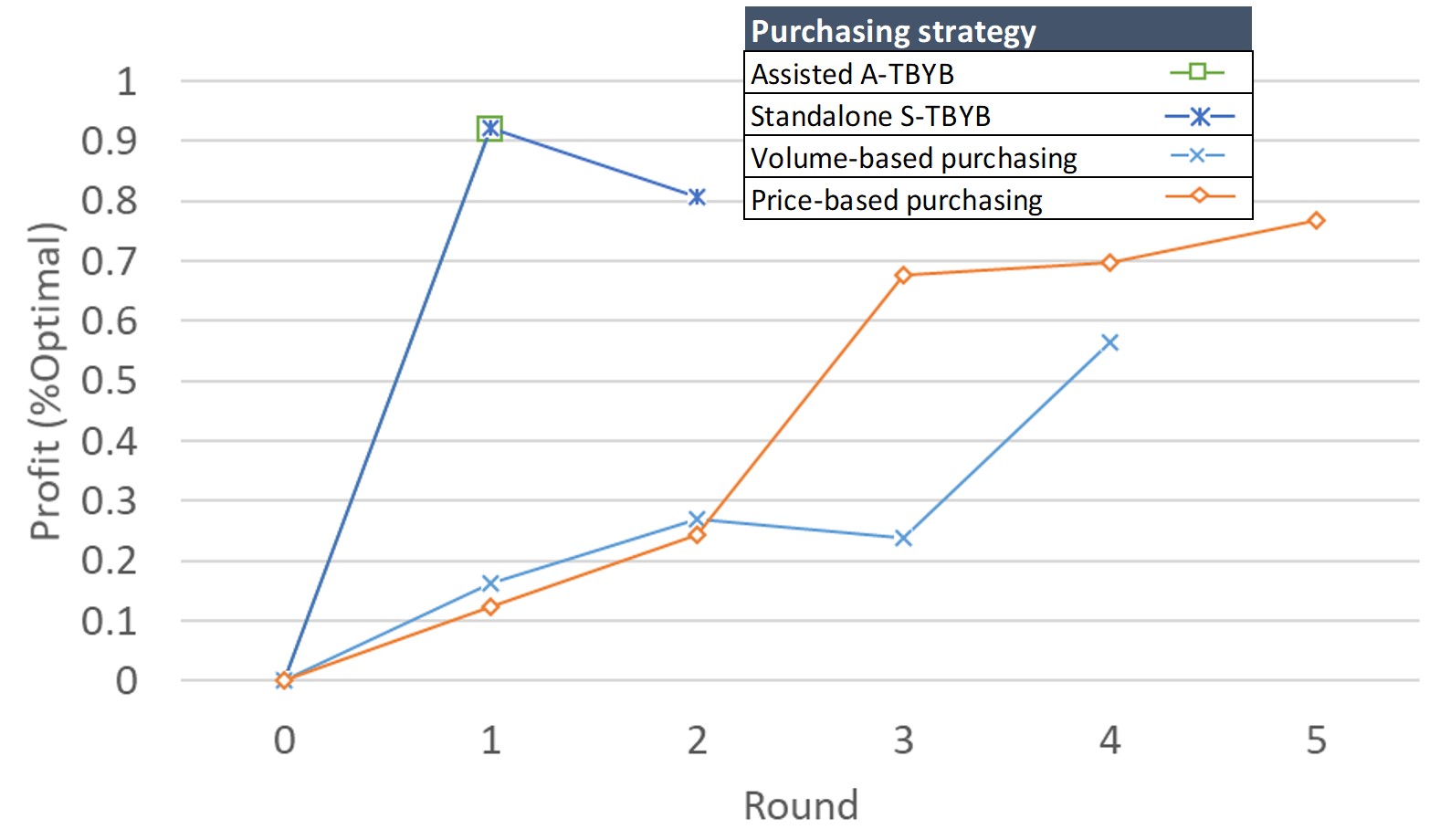}
\caption{Sample purchase sequence for volume-based pricing and TCOD = 3.3}
\label{Fig:PurchasingSequencesD11}
\end{figure}

Figure \ref{Fig:PurchasingSequencesD11} shows an average purchase sequence to understand why TBYB works for volume-related prices. TBYB improves price-based purchasing both by selecting the best datasets, and stopping the purchase process before profit gets diminished. This feature is especially relevant when the offer is wide in comparison to the value they provide (TCOD $>> 1$). Picking datasets based on volume did not improve price-based purchasing in this specific example.

\section{Related work}
\label{sect:RelatedWork}
Different efforts have been done by the research community in order to identify the challenges of existing data marketplaces and define new business models \cite{Fernandez20}. In particular, some ML/AI oriented data marketplace proposals are trying to mechanize in a marketplace the process that some niche data an AI service providers are already bringing to the market. Most AI/ML-oriented theoretical data marketplace platforms leverage a data valuation framework similar to what we propose in this paper \cite{Agarwal19,Chen19}. In general, such marketplaces train buyers’ models in a neutral platform by feeding them with their data, and ask for a price depending on the accuracy (and, thus the value) they provide. They even suggest that marketplaces should return a trained algorithm instead of bulk data. The more you pay, the higher the accuracy you get in any case.

To the best of our knowledge, there is no practical nor commercial implementation of those designs yet, although some digital service providers (SigOpt, Comet.ml), provide algorithm optimization services.

Furthermore, some researchers have studied the dynamics of a data marketplace \cite{Moor19}, and proposed mechanisms that prevent sellers and buyers to postpone their arrival to the marketplace or misreport their costs or values. However, lots of challenges and research issues are still open around data trading and pricing \cite{Pei20, Yang19, Shen16}.

As data provided to buyers usually benefits from combining different sources, it becomes very relevant the problem of how to fairly split the payment for a transaction among all the sources that contributed to the traded data. Existing marketplaces usually calculate this through simple heuristics, such as the data volume or the number of sources involved. However, simple heuristics are not necessarily tied to the utility of data, and consequently could be considered unfair by sellers. To address this challenge, researchers resort to well-known concepts of game theory to split the revenue. Most papers  propose using the Shapley value \cite{Shapley52} for such a task \cite{Agarwal19, Ghorbani19, Paraschiv19}, whereas others propose to use the core \cite{Yan20}. We have used the Shapley value and its approximation algorithms (see \cite{Jia19_2, Castro09, Ghorbani19}) to price datasets according to their value.

\section{Conclusion and future work}
\label{Conclusion}
TBYB was shown to provide near optimal data buyers' profits under a wide range of parameters and data. Used with off-the-shelf AI/ML algorithms, or more complex ones in a sandboxed ``try before you buy'' infrastructure, can make TBYB a practical, high performance alternative to value-unaware purchasing, which is currently the normal in real-world DMs. Helping buyers achieve higher profits would thus help in bootstrapping and growing the currently nascent data marketplace economy.


We are currently working on developing fully functional prototype of TBYB to be used by real users. This will enable us to test arbitrary user-provided algorithms and models. Of course, building a functional DM goes beyond the scope of this paper. It involves additional aspects not covered here such as dynamic data pricing, protection against arbitrage, price discrimination, and lots of engineering, scalability, and security challenges that will be the focus of forthcoming works.



\begin{thebibliography}{00}
\bibitem {Agarwal19} Anish Agarwal, Munther Dahleh, and Tuhin Sarkar. 2019. A Marketplace for Data: An Algorithmic Solution. 701–726. \url{https://doi.org/10.1145/3328526.3329589}
\bibitem {Andres20} Santiago Andrés Azcoitia, Marius Paraschiv and Nikolaos Laoutaris. 2020. Computing the Relative Value of Spatio-Temporal Data in Wholesale and Retail Data Marketplaces. arXiv e-prints, Article arXiv:2002.11193 (May 2020).  arXiv:2002.11193v2 [cs.SI]
\bibitem {Armstrong06} Competition in two-sided markets. The   RAND   Journal   of   Economics 37, 3 (2006). 668–691.
\bibitem {Brovman16} Yuri  M.  Brovman,  Marie  Jacob,  Natraj  Srinivasan,  Stephen  Neola,Daniel Galron, Ryan Snyder, and Paul Wang. 2016.  Optimizing Simi-lar Item Recommendations in a Semi-Structured Marketplace to Maximize Conversion. InProceedings  of  the  10th  ACM  Conference  on  Recommender  Systems(Boston,  Massachusetts,  USA)(RecSys  ’16).  ACM,  New  York,  NY,  USA,  199–202. \url{https://doi.org/10.1145/2959100.2959166}
\bibitem {Castro09} Javier Castro, Daniel Gomez, and Juan Tejada. 2009. Polynomial calculation of the Shapley value based on sampling. Computers and Operations Research 36 (05 2009), 1726–1730. \url{https://doi.org/10.1016/j.cor.2008.04.004}
\bibitem {Chawla19} Shuchi  Chawla,  Shaleen  Deep,  Paraschos  Koutris,  and  Yifeng  Teng. 2019. Revenue  maximization  for  query  pricing. Proceedings  of  the VLDB Endowment13 (09 2019), 1–14. \url{https://doi.org/10.14778/3357377.3357378}
\bibitem {Chen19} Lingjiao  Chen,  Paraschos  Koutris,  and  Arun  Kumar.  2019.   Towards Model-Based  Pricing  for  Machine  Learning  in  a  Data  Marketplace.In Proceedings  of  the  2019  International  Conference  on  Management of  Data(Amsterdam,  Netherlands)(SIGMOD  ’19).  Association  for Computing  Machinery,  New  York,  NY,  USA,  1535–1552.\url{https://doi.org/10.1145/3299869.3300078}
\bibitem {Fernandez20} Michael Franklin. Raul Castro Fernandez, Pranav Subramaniam. 2020. Data Market Platforms: Trading Data Assets to Solve Data Problems. In Proceedings of the VLDB Endowment, Vol. 13. PVLDB, 1933–1947. \url{https://doi.org/10.14778/3407790.3407800}
\bibitem {Ghorbani19} Amirata Ghorbani and James Zou. 2019. Data Shapley: Equitable Valuation of Data for Machine Learning. (04 2019).
\bibitem {Hubert18} Paul Hubert and Giovanni Ricco. 2018. Imperfect information in macroeconomics.
Sciences Po publications info:hdl:2441/7rrg4irjh79. Sciences Po.
\bibitem {Jia19_2} Ruoxi Jia,  David Dao,  Boxin Wang,  Frances Ann Hubis,  Nick Hynes, Nezihe Merve Gurel,  Bo  Li,  Ce  Zhang,  Dawn  Song,  and  Costas  J. Spanos.  2019.   Towards  Efficient  Data  Valuation  Based  on  the  Shapley  Value(Proceedings  of  Machine  Learning  Research,  Vol.  89),  Kamalika  Chaudhuri  and  Masashi  Sugiyama  (Eds.).  PMLR,  1167–1176. \url{http://proceedings.mlr.press/v89/jia19a.html}
\bibitem {Koutris15} Paraschos Koutris, Prasang Upadhyaya, Magdalena Balazinska, Bill Howe, and Dan Suciu. 2015. Query-Based Data Pricing. J. ACM 62, 5, Article 43 (Nov. 2015), 44 pages. \url{https://doi.org/10.1145/2770870}
\bibitem {Moor19} Dmitry Moor. 2019. Data Markets with Dynamic Arrival of Buyers and Sellers. In Proceedings of the 14th Workshop on the Economics of Networks, Systems and Computation (NetEcon ’19). Association for Computing Machinery, New York, NY, USA, Article 3, 6 pages. \url{https://doi.org/10.1145/3338506.3340270}
\bibitem {TaxiTrips19} City of Chicago. 2019 (accessed October, 2020). Taxi Trips. \url{https://data.cityofchicago.org/Transportation/Taxi-Trips/wrvz-psew}
\bibitem {Pei20} Jian Pei. 2020. Data Pricing – From Economics to Data Science. In Proceedings of the 26th ACM SIGKDD International Conference on Knowledge Discovery and Data Mining (KDD ’20). Association for Computing Machinery, New York, NY, USA, 3553–3554. \url{https://doi.org/10.1145/3394486.3406473}
\bibitem {Rochet06} Jean-Charles Rochet and Jean Tirole. 2006. Two-sided markets: a progress report. The RAND Journal of Economics 37,
3 (2006), 645–667.
\bibitem {Paraschiv19} Marius Paraschiv and Nikolaos Laoutaris. 2019. Valuating User Data in a Human-Centric Data Economy. arXiv e-prints, Article arXiv:1909.01137 (Aug 2019). arXiv:cs.SI/1909.01137
\bibitem {Rysman09} Marc Rysman. 2009. The Economics of Two-Sided Markets. Journal of Economic Perspectives 23, 3 (September 2009), 125–43. \url{https://doi.org/10.1257/jep.23.3.125}
\bibitem {Shapley52} Lloyd S. Shapley. 1952. A Value for n-Person Games. (1952). \url{https://www.rand.org/pubs/papers/P0295.html}
\bibitem {Shen16} Yuncheng Shen, Bing Guo, Yan Shen, Xuliang Duan, Xiangqian Dong, and Hong Zhang. 2016. A pricing model for Big Personal Data. Tsinghua Science and Technology 21 (10 2016), 482–490. \url{https://doi.org/10.1109/TST.2016.7590317}
\bibitem {Yan20} Tom Yan and A. Procaccia. 2020. If You Like Shapley Then You’ll Love the Core.
\bibitem {Yang19} Jian Yang, Chongchong Zhao, and Chunxiao Xing. 2019. Big Data Market Optimization Pricing Model Based on Data Quality. Complexity 2019 (04 2019), 1–10. \url{https://doi.org/10.1155/2019/5964068}

\end{thebibliography}
\end{document}